\documentclass{PoS}
      \usepackage{bm}% bold math
\usepackage{multirow}
\usepackage{ctable}
\usepackage{booktabs}
\usepackage{array}
\usepackage{tabularx}
\usepackage{xcolor}
\usepackage{pstricks}
\title{Charmed mesons and leptons \\from semileptonic decays at the LHC}

\ShortTitle{Charmed mesons and non-photonic leptons at the LHC}

\author{\speaker{Rafa{\l} Maciu{\l}a}\\
        Institute of Nuclear Physics PAN, PL-31-342 Cracow, Poland\\
        E-mail: \email{rafal.maciula@ifj.edu.pl}}

\author{Antoni Szczurek\\
        Institute of Nuclear Physics PAN, PL-31-342 Cracow, Poland\\
        University of Rzeszow, PL-35-959 Rzeszow, Poland\\
        E-mail: \email{antoni.szczurek@ifj.edu.pl}}

\abstract{We discuss production of charmed mesons as well as electrons/muons from semileptonic decays of charm and bottom mesons in proton-proton collisions at the LHC. The cross section for inclusive production of $c \bar c$ and $b \bar b$ pairs is calculated in the framework of the $k_{\perp}$-factorization approach. Here, the KMR and Jung CCFM unintegrated gluon distribution functions are used. Theoretical uncertainties of the model related to the choice of renormalization and factorization scales as well as due to the quark mass are also discussed. The hadronization of charm and bottom quarks is included within the fragmentation functions technique. Inclusive differential distributions in transverse momentum of charmed mesons are presented and compared to recent results of the ALICE collaboration. Furthermore, we also consider production of different $D \overline D$ pairs in unique kinematics of forward rapidities of the LHCb experiment. Kinematical correlations in azimuthal angle $\varphi_{D\overline D}$ and invariant mass $M_{D \overline D}$ distributions are presented and compared to LHCb data. Furthermore, the semileptonic decays of charm and bottom mesons are done with the help of decay functions found by fitting recent semileptonic data obtained by the CLEO and BABAR collaborations. Inclusive differential distributions in transverse momentum of leptons for several kinematical regions are presented and compared to recent results of the ALICE and CMS collaborations.}

\FullConference{XXI International Workshop on Deep-Inelastic Scattering and Related Subject -DIS2013,\\
		22-26 April 2013\\
		Marseilles,France}

\begin{document}

\section{Heavy quarks production within the \bm{$k_t$}-factorization approach}

In the studies of heavy quark hadroproduction the main efforts usually concentrate on inclusive distributions. The transverse momentum distribution of charmed mesons or heavy flavoured leptons are the best examples. From the theoretical point of view, the improved schemes of standard NLO collinear framework e.g. FONLL \cite{FONLL}, are states of art in this respect. These approaches cannot be, however, used  when transverse
momenta of outgoing quark and antiquark are not equal. This means in practice 
that they cannot be used for studies of correlation observables.

The $k_t$-factorization is the approach which is very efficient in studies of kinematical correlations and can be used to describe many high-energy processes (see e.g. Ref.~\cite{MS2013} and references therein). In this sense it is an alternative to standard collinear-factorization approach. If one allows for transverse momenta of incident partons the cross section for the production of heavy quark $Q\overline Q$ pairs in proton-proton collisions can be written as:
\begin{eqnarray}
\frac{d \sigma(p p \to Q \overline Q \, X)}{d y_1 d y_2 d^2 p_{1t} d^2 p_{2t}} 
&& = \frac{1}{16 \pi^2 {\hat s}^2} \int \frac{d^2 k_{1t}}{\pi} \frac{d^2 k_{2t}}{\pi} \overline{|{\cal M}^{off}_{g^{*}g^{*} \to Q \overline Q}|^2} \nonumber \\
&& \times \;\; \delta^2 \left( \vec{k}_{1t} + \vec{k}_{2t} - \vec{p}_{1t} - \vec{p}_{2t}
\right)
{\cal F}_g(x_1,k_{1t}^2,\mu^2) {\cal F}_g(x_2,k_{2t}^2,\mu^2).
\end{eqnarray}
The main ingredients in the formula are off-shell matrix elements for $g^{*}g^{*} \rightarrow Q \overline{Q}$ subprocess
and unintegrated gluon distributions (UGDF). The relevent matrix
elements are known and can be found e.g. in Ref.~\cite{CCH91}. 
The unintegrated gluon distributions are functions of
longitudinal momentum fraction $x_1$ or $x_2$ of gluon with respect to its parent nucleon and of gluon transverse momenta $k_{t}$.
Some of them depend in addition on the
factorization scale $\mu$.

Various UGDFs have been discussed in the literature
(see Ref.~\cite{MS2013} and references therein). In contrast to the collinear gluon distributions (PDFs) they differ
considerably among themselves. One may expect that they will lead to different
production rates of $c \bar c$ and $b \bar b$ pairs at the LHC. Since the production of charm and bottom quarks
is known to be dominated by the gluon-gluon fusion, the heavy flavours production at the LHC
can be used to verify the quite different models of UGDFs. It has been shown in Ref.~\cite{MS2013} that in the case of charm production at $\sqrt{s} = 7$ TeV and at forward rapidities $|y| > 3$, one starts to probe $x$-values smaller than 10$^{-4}$. This is a new situation compared to earlier measurements at RHIC or Tevatron. The unintegrated gluon distributions (UGDFs) as well as standard collinear ones (PDFs) were not tested so far in this region.

\section{Inclusive open charmed mesons spectra}

The hadronization of heavy quarks is usually done
with the help of fragmentation functions. The inclusive distributions of
open heavy mesons can be obtained through a convolution of inclusive distributions
of heavy quarks/antiquarks and $Q \to M$ fragmentation functions:
\begin{equation}
\frac{d \sigma(pp \rightarrow M \overline{M} \, X)}{d y_M d^2 p_{t,M}} \approx
\int_0^1 \frac{dz}{z^2} D_{Q \to M}(z)
\frac{d \sigma(pp \rightarrow Q \overline{Q} \, X)}{d y_Q d^2 p_{t,Q}}
\Bigg\vert_{y_Q = y_M \atop p_{t,Q} = p_{t,M}/z}
 \; ,
\label{Q_to_h}
\end{equation}
where $p_{t,Q} = \frac{p_{t,M}}{z}$ and $z$ is the fraction of
longitudinal momentum of heavy quark carried by meson.
We have made typical approximation assuming that $y_{Q}$ is
unchanged in the fragmentation process.

As a default set in our calculations we use standard Peterson model of fragmentation function \cite{Peterson} with
the parameters $\varepsilon_{c} = 0.02$ and $\varepsilon_{b} = 0.001$. These values are sligthly smaller than those extracted by ZEUS and H1 analyses, however, they are consistent with the fragmentation scheme applied in the FONLL framework, where also rather harder functions (or smaller $\varepsilon$'s) are used. This issue together with effects of applying other fragmentation functions from the literature is carefully studied in Ref.~\cite{MS2013}

Recently, the ALICE collaboration, has measured inclusive 
distributions (mainly transverse momentum distributions) of different charmed mesons \cite{ALICEincD}.

\begin{figure}[!h]
\centering
\begin{minipage}{0.4\textwidth}
 \centerline{\includegraphics[width=1.0\textwidth]{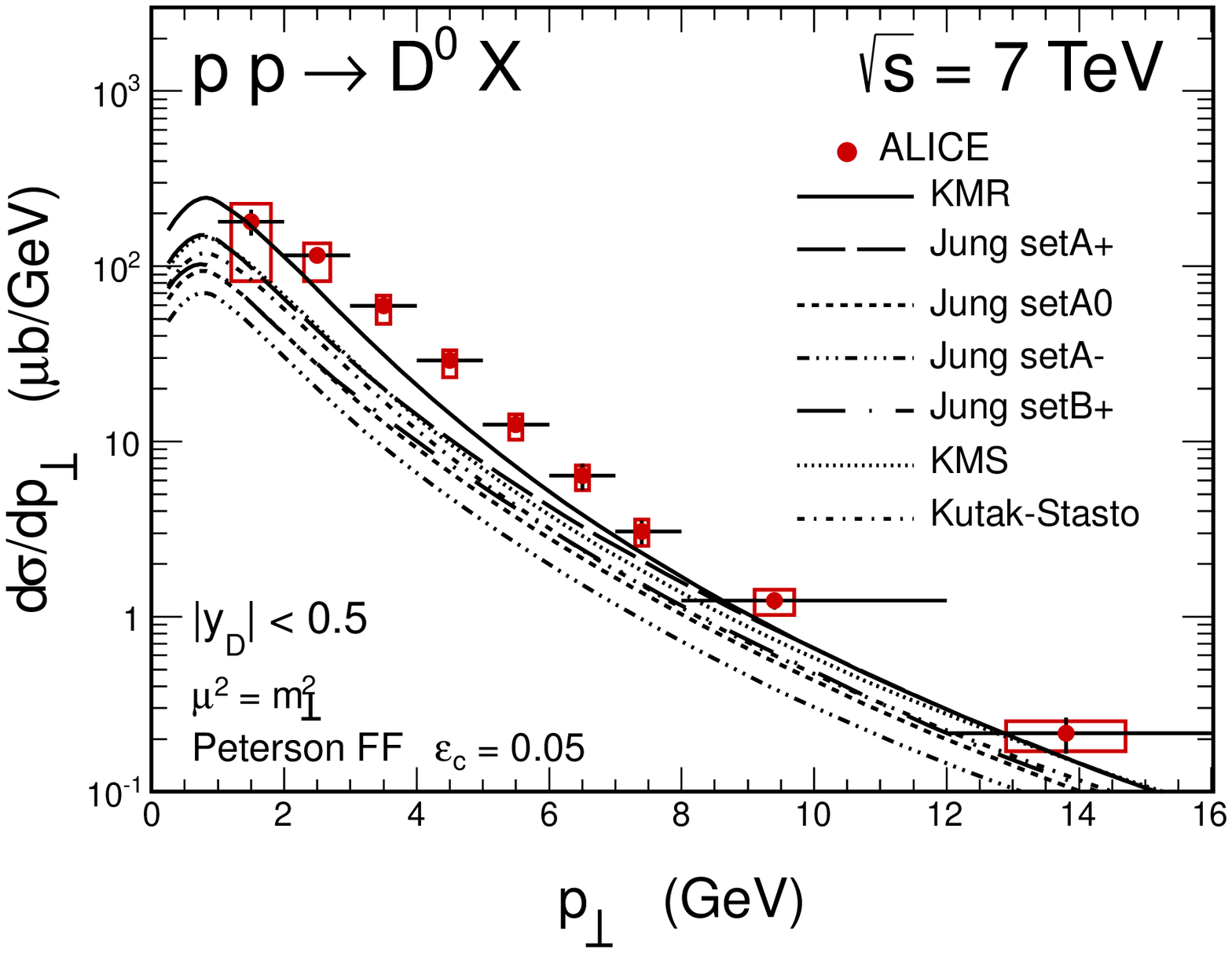}}
\end{minipage}
\hspace{0.5cm}
\begin{minipage}{0.4\textwidth}
 \centerline{\includegraphics[width=1.0\textwidth]{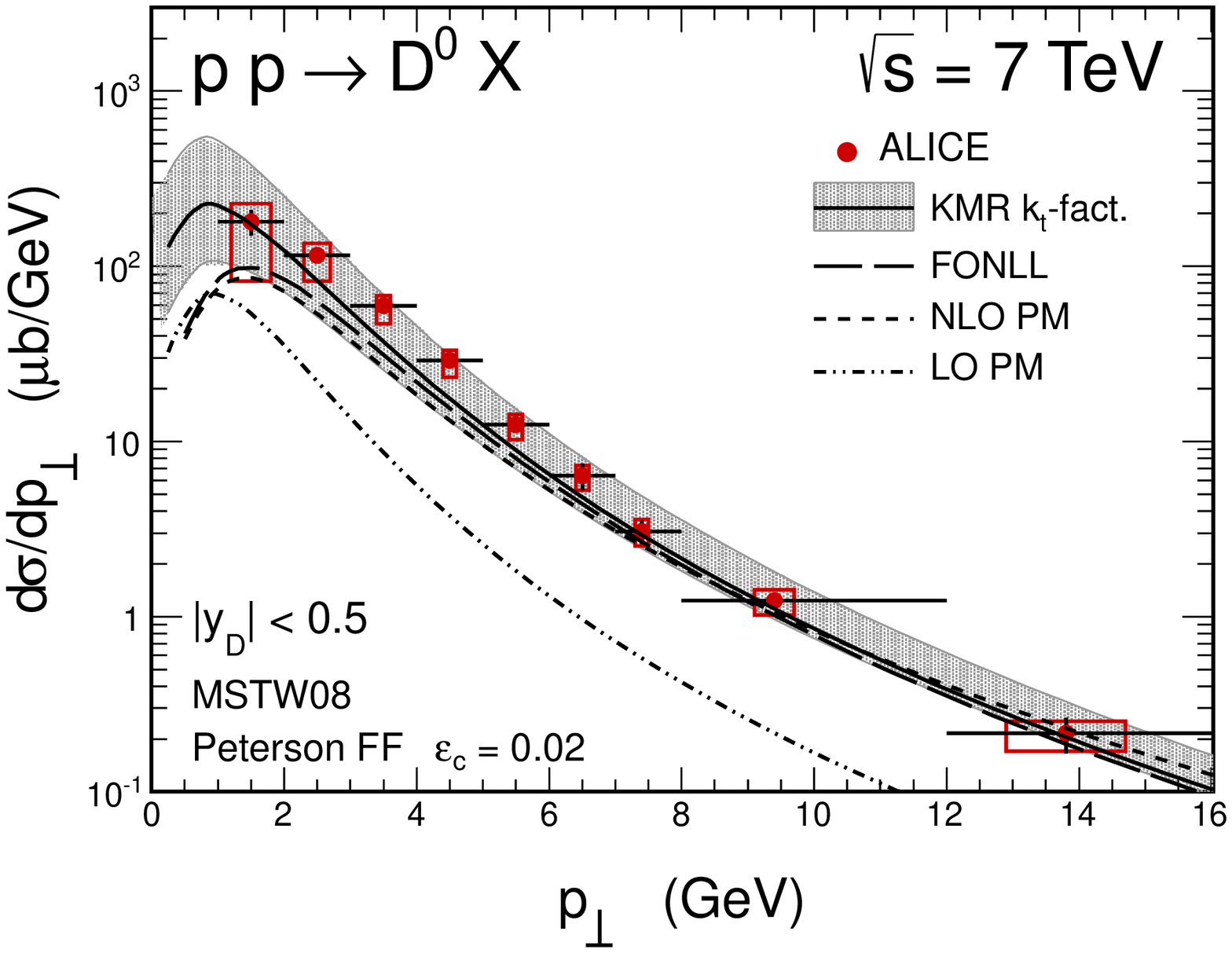}}
\end{minipage}
   \caption{
\small Inclusive $D$ meson transverse momentum distributons compared with the ALICE data for different UGDFs (left panel) and combined with the results of the collinear calculations (right panel). The shaded uncertainty band corresponds
to the uncertainties of our predictons due to
factorization/renormalization scale and those related with quark mass uncertainties. 
}
 \label{fig:pt-alice-D-1}
\end{figure}

In the left panel of Fig.~\ref{fig:pt-alice-D-1} we show $p_t$ distribution
of $D^0$ mesons calculated with different UGDFs
known from the literature. Most of the existing distributions fail
to describe the ALICE data. Only the KMR UGDF provides reasonably good description
of the measured distributions.

The right panel of Fig.~\ref{fig:pt-alice-D-1} presents a comparison
of the results of the $k_t$-factorization approach with the KMR UGDF to those obtained within
LO, NLO PM and FONLL frameworks. The cross sections obtained within
leading-order collinear approximation (LO PM) are much smaller, in particular
for larger $p_t$'s. However, a good agreement between the rest of plotted distributions
can be observed. The result of the KMR UGDF is consistent with the FONLL predictions in a wide range of transverse momenta.
Only at $p_t$'s less than $3$ GeV some differences appear. This is the region when transverse momenta of incoming gluons play important role and a detailed treatment of the non-perturbative $k_t$ region may lead to different behaviour
of the cross sections in this regime. 

%-----------------------------------------------------------------------------
%------------------------------------------------------------------------------

\section{Kinematical correlations of \bm{$D \overline D$} pairs}

Most of the calculations in the literature concentrates on single meson distributions.
We wish to focus now on correlation observables for $D$ and $\bar{D}$ mesons.
In order to calculate correlation observables for two mesons we follow here, similar as in the single meson case,
the fragmentation function technique for hadronization process, but now a multidimensional distribution
containing all relevant kinematical informations about both quark and antiquark is convoluted with respective
fragmentation functions simultaneously. As a result of the hadronization one obtains corresponding two-meson 
multidimensional distribution. 

%-----------------------------------------------------------------------------
\begin{figure}[!h]
\centering
\begin{minipage}{0.4\textwidth}
 \centerline{\includegraphics[width=1.0\textwidth]{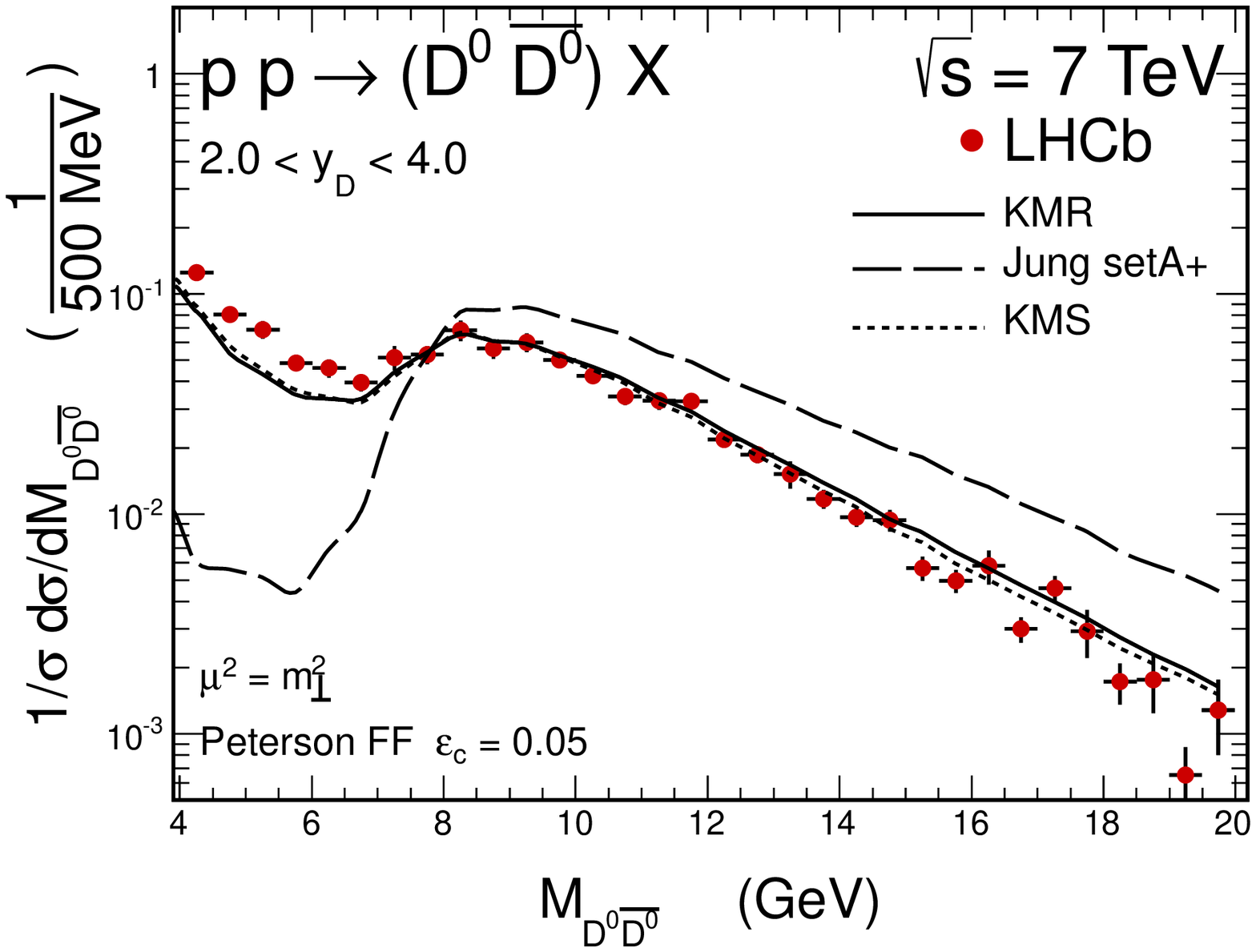}}
\end{minipage}
\hspace{0.5cm}
\begin{minipage}{0.4\textwidth}
 \centerline{\includegraphics[width=1.0\textwidth]{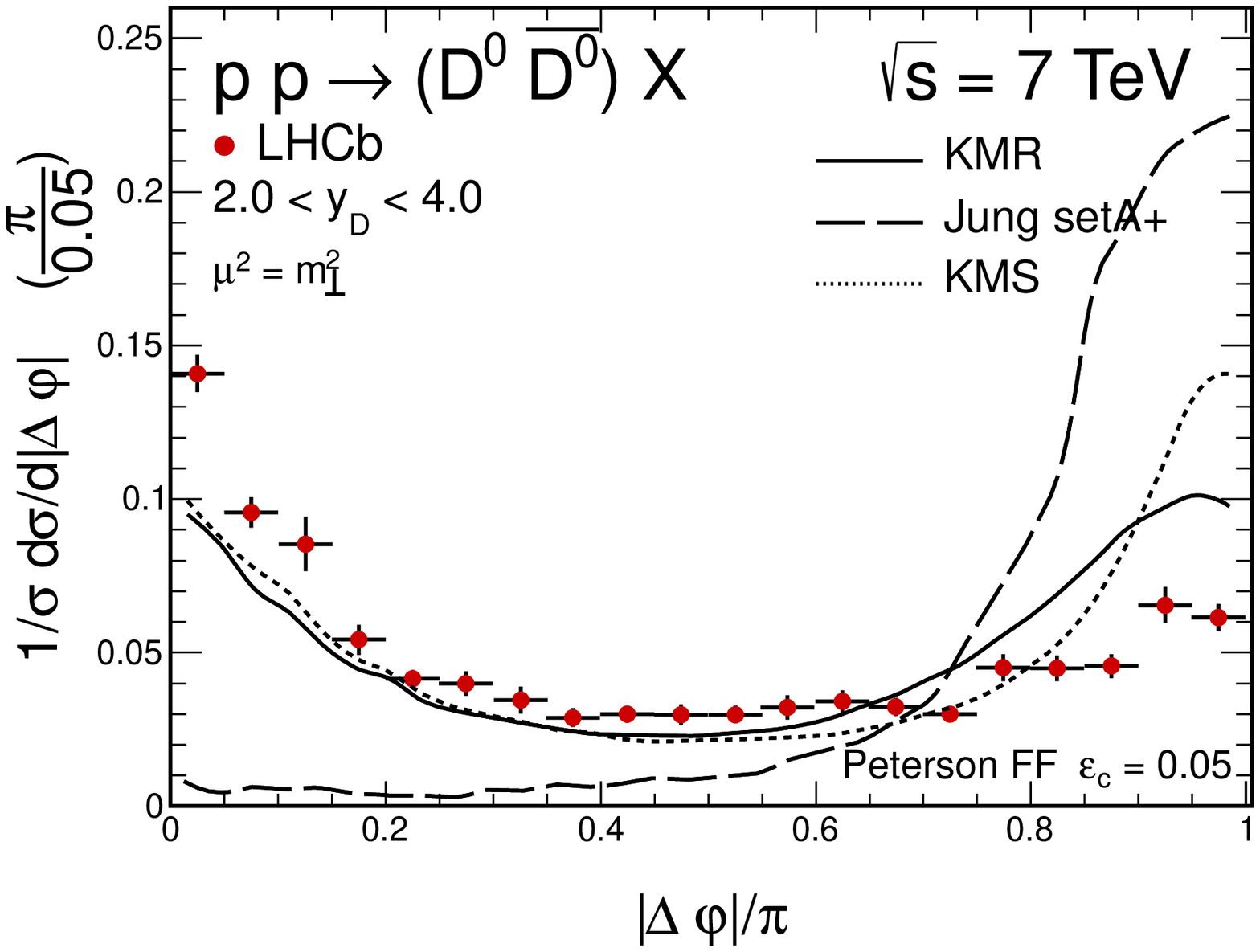}}
\end{minipage}
   \caption{
\small Invariant mass distribution of the $D^0 \overline{D^0}$ system (left) and distribution in relative azimuthal
 angle between $D^0$ and $\overline{D^0}$ for different UGDFs, compared with the LHCb data.
}
 \label{fig:minv-lhcb-DDbar-2}
\end{figure}
%------------------------------------------------------------------------------

The LHCb collaboration presented also distribution in the $D^0 \bar D^0$
invariant mass $M_{D^0 \bar D^0}$ \cite{LHCb-DPS-2012}. In the left panel of Fig.~\ref{fig:minv-lhcb-DDbar-2} we show the
corresponding theoretical result for different UGDFs. Both, the KMR and KMS UGDFs provide
right shape of the distribution. The dip at small invariant masses
is due to specific LHCb cuts on kinematical variables.

In turn, in the right panel of Fig.~\ref{fig:minv-lhcb-DDbar-2} we discuss distribution in azimuthal
angle between the $D^0$ and $\bar D^0$ mesons $\varphi_{D^0 \bar D^0}$. Again the KMR and KMS
distributions give quite reasonable description of the shape of the
measured distribution. Both of them, give an enhancement of the cross section
at $\phi_{D \bar D} \sim$ 0. This is due to the fact that these
approaches include effectively gluon splitting contribution, not included
in the case of the Jung UGDFs. However, still even with the KMR UGDF, one can observe some small missing strenght at small angles. It may suggest that within the KMR model the gluon splitting contribution is not fully
included.

In principle, we get good agreement with the LHCb data not only in the shapes of the correlation distributions but also when comparing integrated cross sections (see Ref.~\cite{MS2013}).

\section{Semileptonic decays of $D$ and $B$ mesons}

The differential cross sections for the so-called non-photonic leptons
which come from semileptonic decays of open charm and bottom mesons
can be obtained by a convolution of the meson-level cross sections with
the semileptonic decay functions \cite{LMS2008}.

In principle the semileptonic decay functions can be calculated, however, it involves
extra source of uncertainties and is not an easy task.
In our approach we follow more pragmatic way and we use decay functions
which are fitted to recent semileptonic $D$ and $B$ data. 
These functions after renormalizing to experimental branching fractions are used to generate electrons/positrons 
in the rest frame of the decaying $D$ and $B$ mesons in a Monte Carlo approach. 

\begin{figure}[!h]
\centering
\begin{minipage}{0.4\textwidth}
 \centerline{\includegraphics[width=1.0\textwidth]{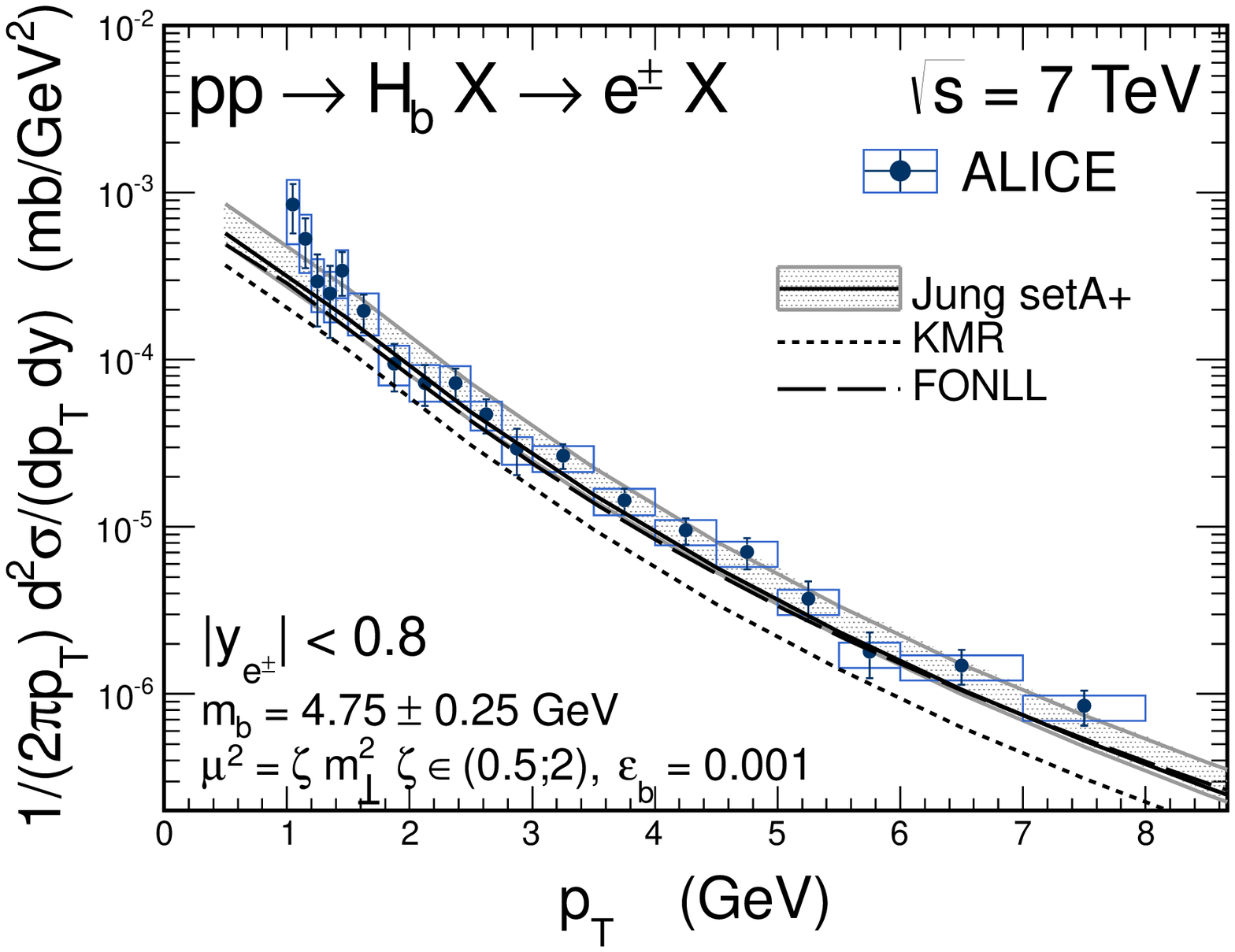}}
\end{minipage}
\hspace{0.5cm}
\begin{minipage}{0.4\textwidth}
 \centerline{\includegraphics[width=1.0\textwidth]{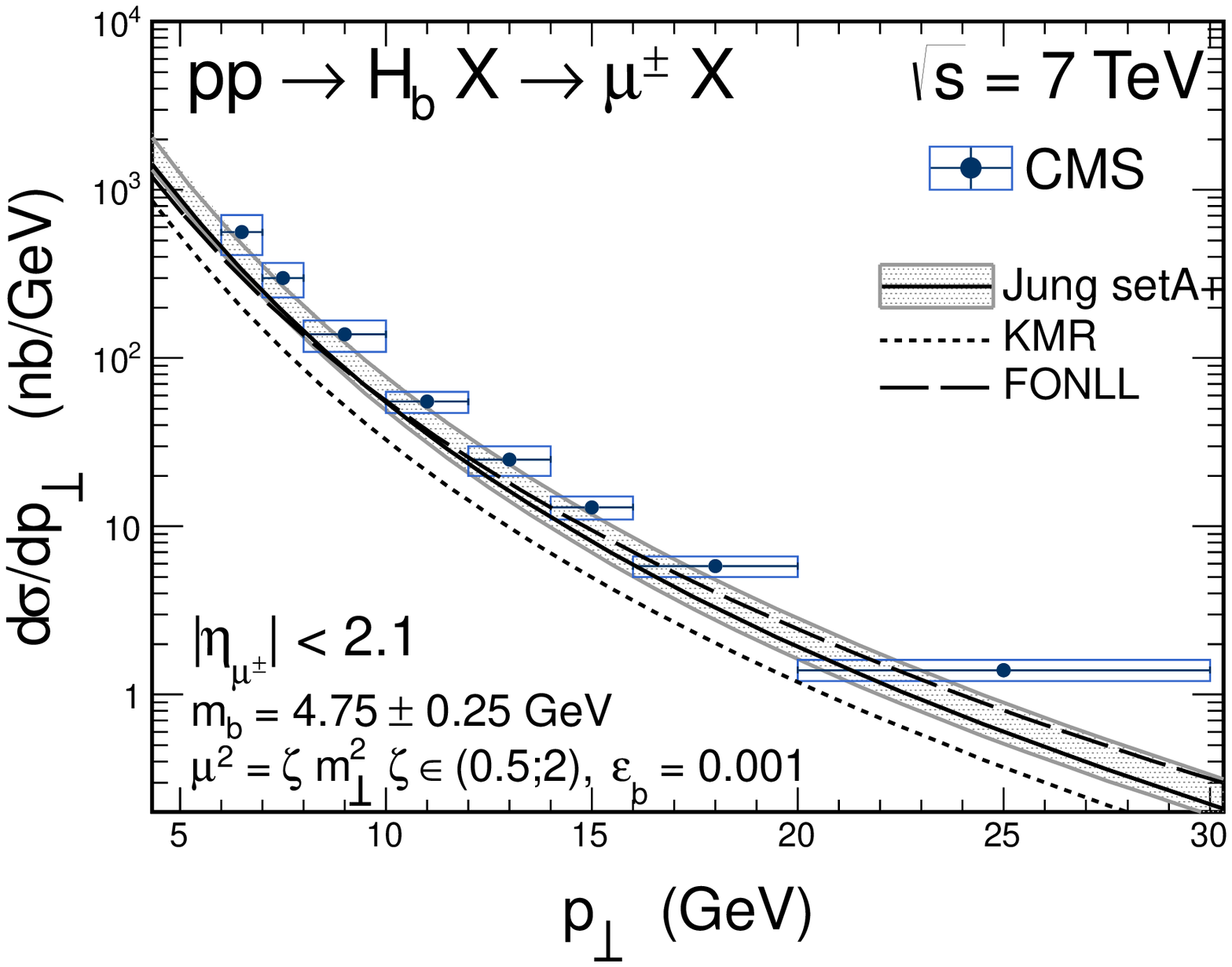}}
\end{minipage}
   \caption{
\small Transverse momentum distributions of leptons from semileptonic decays of bottom mesons for ALICE (left) and CMS (right) obtained with different UGDFs and compared with results of FONLL. The shaded uncertainty bands correspond
to the uncertainties of our predictons due to
factorization/renormalization scale and those related with quark mass.
}
 \label{fig:npe-bottom-only}
\end{figure}

In Fig.~\ref{fig:npe-bottom-only} we present the results of our calculations compared to the measured
transverse momentum spectra of the leptons coming from the decay of open bottom mesons \cite{bottom-lep}.
Two different models of UGDFs which were previously tested at the meson-level are used.
The ALICE and CMS data are well described by the Jung setA$+$ UGDFSa
which by chance coincides with
the FONLL predictions. The KMR UGDF in the case of bottom flavoured leptons underestimates the data points. 

\begin{figure}[!h]
\centering
\begin{minipage}{0.4\textwidth}
 \centerline{\includegraphics[width=1.0\textwidth]{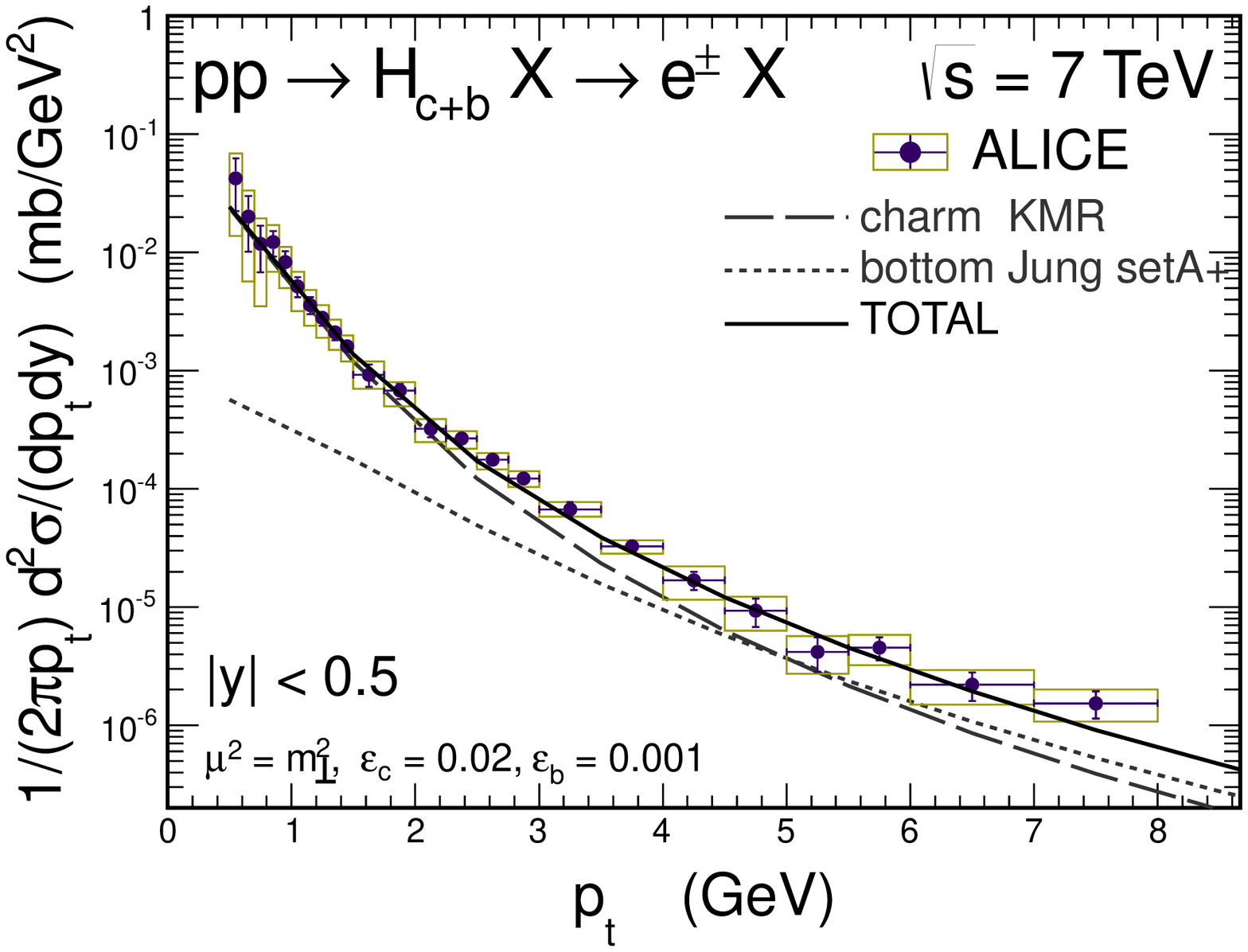}}
\end{minipage}
\hspace{0.5cm}
\begin{minipage}{0.4\textwidth}
 \centerline{\includegraphics[width=1.0\textwidth]{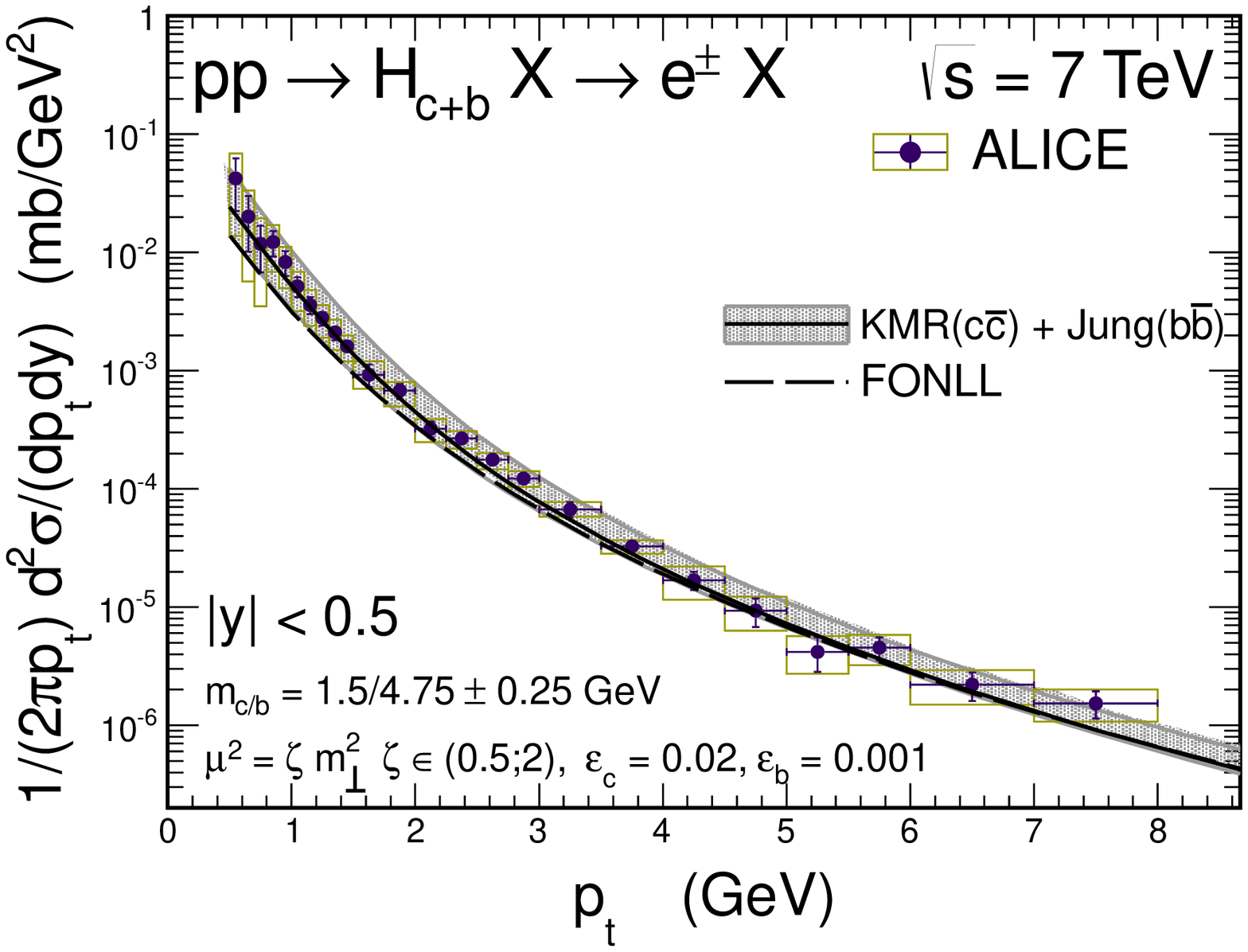}}
\end{minipage}
   \caption{
\small Transverse momentum distributions of leptons from semileptonic
decays of charm and bottom mesons with the ALICE data. The left panel
shows separately the charm and bottom 
components and the right panel represents theoretical uncertainties of our calculations.
}
 \label{fig:npe-total}
\end{figure}

Thus, considering the ALICE summed lepton spectra \cite{sum-lep} with charm and bottom contributions we combine the calculations by using the KMR UGDF for charm and Jung setA$+$ for bottom cross sections. Such recipe gives excelent
description of the data and provides better consistency than FONLL at low lepton $p_t$'s (see Fig.~\ref{fig:npe-total}). This analysis shows that there is no unique model of UGDFs which correctly describes both charm and bottom cross sections at the LHC.

%-----------------------------------------------------------------------------
%------------------------------------------------------------------------------
\vspace{-2mm}
%-----------------------------------------------------------------------------
%------------------------------------------------------------------------------
\vspace{-1mm}
\noindent
{\bf Acknowledgments}\\
This work is supported in part by the Polish
Grants DEC-2011/01/B/ST2/04535 and N202 237040.

\end{document}